\documentclass[lettersize,journal]{IEEEtran}
\usepackage{amsmath,amsfonts}
\usepackage{algorithmic}
\usepackage{algorithm}
\usepackage{array}
\usepackage[caption=false,font=normalsize,labelfont=sf,textfont=sf]{subfig}
\usepackage{textcomp}
\usepackage{stfloats}
\usepackage{url}
\usepackage{verbatim}
\usepackage{graphicx}
\usepackage{multirow}
\usepackage{cite}
\usepackage{soul}
\DeclareMathOperator*{\argmax}{arg\,max}
\DeclareMathOperator*{\argmin}{arg\,min}

\begin{document}

\title{Deep Temporal Sequence Classification and Mathematical Modeling for Cell Tracking in Dense 3D Microscopy Videos of Bacterial Biofilms}

\author{Tanjin Taher Toma, Yibo Wang, Andreas Gahlmann, Scott T. Acton
\thanks{Tanjin Taher Toma is with the Department of Electrical and Computer Engineering, University of Virginia, Charlottesville, VA 22904 USA
(e-mail: tt4ua@virginia.edu).}
\thanks{Yibo Wang is with the Department of Chemistry, University of Virginia, Charlottesville, VA 22904 USA
(e-mail: yw9et@virginia.edu).}
\thanks{Andreas Gahlmann is with the Department of Chemistry, University of Virginia, Charlottesville, VA 22904 USA and the Department of Molecular Physiology and Biological Physics, University of Virginia, Charlottesville, VA 22903 USA
(e-mail: ag5vu@virginia.edu).}
\thanks{Scott T. Acton is with the Department of Electrical and Computer Engineering, University of Virginia, Charlottesville, VA 22904 USA
(e-mail: acton@virginia.edu).}
}



\maketitle

\begin{abstract}
Automatic cell tracking in dense environments is plagued by inaccurate correspondences and misidentification of parent-offspring relationships. In this paper, we introduce a novel cell tracking algorithm named \textit{DenseTrack}, which integrates deep learning with mathematical model-based strategies to effectively establish correspondences between consecutive frames and detect cell division events in crowded scenarios. 
We formulate the cell tracking problem as a deep learning-based temporal sequence classification task followed by solving a constrained one-to-one matching optimization problem exploiting the classifier's confidence scores. Additionally, we present an eigendecomposition-based cell division detection strategy that leverages knowledge of cellular geometry. The performance of the proposed approach has been evaluated by tracking densely packed cells in 3D time-lapse image sequences of bacterial biofilm development. The experimental results on simulated as well as experimental fluorescence image sequences suggest that the proposed tracking method achieves superior performance in terms of both qualitative and quantitative evaluation measures compared to recent state-of-the-art cell tracking approaches.

\end{abstract}


\section{Introduction}
\IEEEPARstart{C}{ell} tracking in time-lapse microscopy image sequences is a challenging multi-object tracking task that is essential for research focusing on the behaviors of individual cells in a population. Because large numbers of cells need to be tracked to make statistically significant conclusions, accurate and robust automated tracking approaches are required. Automated tracking involves identifying and linking instances of the same biological cell and perhaps their offspring in consecutive frames of an image sequence. Accurate reconstructions of cell trajectories enables researchers to extract biologically and biophysically relevant parameters, such as cell growth and division rate, cell adhesion and dispersal frequencies, death rate, as well as changes in cellular motion patterns. All these observables can provide quantitative insights into how population behaviors emerge from the underlying behaviors of individual cells~\cite{ulman2017objective,mavska2014benchmark}. The cell tracking problem often becomes challenging to solve in the presence of high cell density, fast motion, and frequent division events. 

There are two main categories of automatic cell tracking methods: tracking-by-contour evolution and tracking-by-detection. The contour evolution-based methods involve finding the object contour in the current frame given an initial contour from the previous frame~\cite{zimmer2002segmentation,ray2002tracking,li2008cell,dzyubachyk2010advanced}. Contour evolution-based approaches solve the segmentation and tracking tasks simultaneously by solving an iterative PDE-based energy functional. In contrast, the tracking-by-detection approach separates the segmentation and tracking task by first performing the segmentation of the individual instances in all the frames and then establishing the temporal associations between the segmented cells of consecutive frames~\cite{kachouie2006probabilistic,rapoport2011novel,rempfler2018tracing}. While tracking-by-contour evolution is effective in certain scenarios, such as when morphological changes of cells are imaged at high magnification, it suffers in situations with low frame rates, high cell density, high motility, and frequent cell divisions. This is due to the underlying assumption of unambiguous spatiotemporal overlap between the corresponding cell regions~\cite{ulman2017objective,mavska2014benchmark}. Tracking-by-detection methods are more effective in such scenarios, and their reduced computational complexity has further led to their widespread adoption for tracking large numbers of cells over longer time periods~\cite{magnusson2014global,bise2011reliable}. In this paper, we focus on tracking-by-detection and present an algorithm to effectively track crowded cells over time in 4D (3D space plus time) data.

Over the years, numerous tracking-by-detection approaches have been proposed. The simplest methods use basic nearest-neighbor techniques to match cells between frames based on features such as intensity distribution, morphology, and size~\cite{dewan2011tracking,boukari2018automated}. More complex features, such as features of the cell's neighborhood~\cite{li2009multiple} or features derived from a graph structure~\cite{narayanaswamy2012multi}, have also been exploited. However, nearest-neighbor methods that rely on a  distance or similarity function are not effective for establishing correspondence in dense cell tracking scenarios, as they often lead to incorrect associations due to sub-optimal user-defined distance or similarity measures~\cite{chenouard2014objective,jaqaman2008robust}. There also exist graph-based tracking approaches where cells are represented as nodes in a graph, and association hypotheses are represented as edges linking the nodes~\cite{padfield2011coupled,kausler2012discrete,liu2019deepseed,schiegg2015graphical}. Such structures allow the tracking problem to be formulated as a graph-matching problem. However, the underlying problem formulation of graph-based tracking methods typically entails solving an optimization problem with numerous regularization terms, which poses challenges in tuning many hyperparameters.

Furthermore, probabilistic approaches for correspondence finding have also been proposed. These include joint probabilistic data association (JPDA) \cite{godinez2014tracking,rezatofighi2012application} and
multiple hypothesis-based tracking (MHT) \cite{chenouard2013multiple,coraluppi2011multi,liang2014novel}. The classical Kalman filter or its probabilistic variants have also been used to predict the position of the cells in the next frame~\cite{liu2017plant,ong2010tracking}. While these traditional methods have demonstrated effectiveness in many applications, they often rely on simplistic assumptions about cell behavior. For example, they may depend on a specific cell motion model or the selection of a particular probability distribution to represent the likelihood of object appearance and disappearance within the field of view. These assumptions do not necessarily hold true in all scenarios~\cite{spilger2020recurrent}. Most importantly, these classical approaches are fully based on fixed models and hence cannot leverage the advantages of learning representative information from a training dataset.

The utilization of deep learning techniques in cell tracking has typically been limited by the unavailability of ground truth annotations for time-lapse image sequences, in particular for 3D images. Several deep learning-based methods have been developed for cell tracking. One such approach models cell tracking as an edge classification problem in a direct graph using a graph neural network~\cite{ben2022graph}. While estimating the entire set of cell trajectories at once with a graph neural network seems efficient, it can lead to numerous incorrect associations, particularly in dense or long image sequences, as many edges need to be classified simultaneously. Another cell tracking approach employs two separate U-Nets for cell likelihood detection and motion estimation~\cite{hayashida2019cell}. Although the motion estimation strategy can be useful for tracking high-motility cells, simply relying on likelihood detection may not be as effective as segmenting all the cells prior to tracking in the case of dense neighborhoods. Other recent approaches that rely on extensive training data and high computational resources include a deep reinforcement learning method~\cite{wang2020deep} and a pipeline of Siamese networks~\cite{panteli2020siamese}, both of which generally depend on large training datasets for optimal performance~\cite{cruciata2021use,ondravsovivc2021siamese}. Furthermore, authors in \cite{loffler2022embedtrack} presented a single convolutional neural network for simultaneous cell segmentation and tracking by predicting cellular embeddings and clustering bandwidths. Its effectiveness, though, has only been demonstrated in the context of cell tracking within 2D image sequences.

A common drawback of all these above-mentioned deep learning-based tracking approaches is that they do not explicitly enforce one-to-one matching between successive frames, the lack of which can lead to erroneous one-to-many associations. Additionally, these methods do not incorporate temporal history to predict associations in the next frame, which can be necessary when a cell is poorly imaged or segmented in some frames but better detected in neighboring frames. Furthermore, some of these approaches have been developed only for 2D cell tracking.

In our proposed approach, we address these limitations of existing classical and deep learning-based methods and present an effective solution that combines a robust deep learning strategy with mathematical modeling for 3D cell tracking in dense environments. Additionally, we overcome the challenge of lacking ground truth annotations for cell tracking by generating training annotations automatically. We achieve this by simulating synthetic biofilm image sequences using a dedicated simulation framework~\cite{toma2022realistic}, which are then used to train the deep learning network in our tracking system.

The key contributions of our method are outlined below: 
\begin{itemize}
    \item For frame-by-frame instance matching, our approach estimates association scores for potential matches in the next frame by conducting a deep learning-based temporal sequence classification task. This enables us to learn the association task through a data-driven approach rather than relying on a fixed classical model. Additionally, such an association score estimation network is trainable with limited training data compared to existing deep learning-based cell tracking methods, as the underlying problem being addressed is a straightforward binary classification task of predicting correct versus incorrect associations given a sequence of spatiotemporal features of instances.
    \item We leverage near-temporal history in our spatiotemporal feature representation to estimate the association scores, instead of relying solely on features from the current and the next frame.
    \item We enforce one-to-one matching between successive frames by solving an optimization problem that utilizes association scores estimated by the classifier. This additional matching step, unlike existing deep learning-based tracking methods, reduces matching errors in high cell density environments, such as those often seen in bacterial biofilms.
    \item For detecting cell division events in bacterial biofilms, we present a novel strategy based on the eigendecomposition of unmatched instances in the next frame. This approach effectively identifies the offspring and their parent instance, even in scenarios where there is minimal overlap between the parent and daughter instances in the next frame due to cell motion.
\end{itemize}

\begin{figure*}[!t]
\centering
\subfloat[]{\includegraphics[width=6.5in]{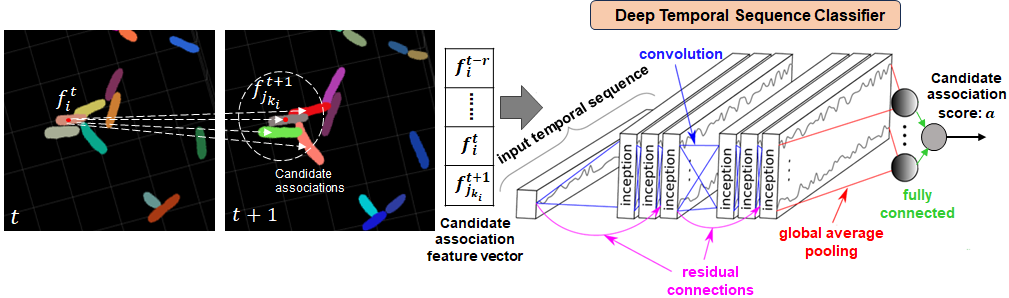}%
\label{tracking_block_diagram_1}}
\hfil 
\subfloat[]{\includegraphics[width=3.0in]{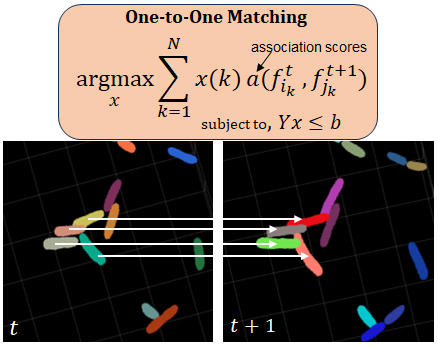}%
\label{tracking_block_diagram_2}}
\hfil
\\
\subfloat[]{\includegraphics[width=3.0in]{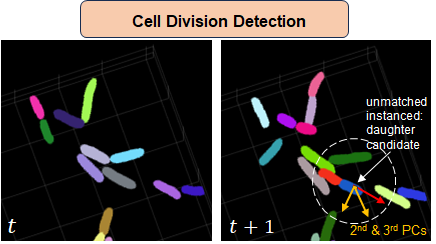}%
\label{tracking_block_diagram_3}}
\caption{Overview of the proposed tracking approach \textit{DenseTrack}. In (a) and (b), we depict our frame-by-frame matching technique, which entails calculating deep learning-based association scores and integrating them into one-to-one matching optimization. (c) illustrates the detection of a cell division event by identifying the neighboring instance with the minimum projection along the $2^{nd}$ and $3^{rd}$ principal components of the unmatched instance in frame $t+1$.}
\label{fig_tracking_block_diagram}
\end{figure*}

\section{Problem Statement}
Let us consider an image sequence, denoted by $\boldsymbol{S}=\{\boldsymbol{F}^{t}\}_{t=1}^{T}$, which comprises $T$ frames. Let $L$ be the number of biological cells present in this sequence. The cell tracking problem can be stated as follows, (1) determine the trajectory of each biological cell and (2) identify the parent of each biological cell in cases where cell existence is due to cell division. For each biological cell, we need to calculate a set of information represented by $\mathcal{\boldsymbol{T}}_{l}=\{t_{init}^{l},t_{fin}^{l},\boldsymbol{C}^{l},P(l)\}$. Here $t_{init}^{l}$ and $t_{fin}^{l}$ refer to the first and last time points in which the $l^{th}$ cell appears in the sequence, respectively. $\boldsymbol{C}^{l}$ represents the set of coordinates of the $l^{th}$ cell from the first frame $t_{init}^{l}$ to the last frame $t_{fin}^{l}$. Finally, $P(l)$ is a function that identifies the parent cell of the $l^{th}$ cell, where $P(l)=l^{'}$ if $l^{'}$ is the parent of cell $l$, and $P(l)=0$ if the cell appearance is not due to cell division. The objective of cell tracking is to obtain the set $\{\mathcal{\boldsymbol{T}}_{1},...,\mathcal{\boldsymbol{T}}_{L}\}$.
\section{Proposed Solution}
\label{sec:tracking_theory}
To solve the problem, our method involves initially matching cell instances across consecutive frames, followed by the detection of division events and the establishment of complete trajectories. An overview of the proposed approach, called \textit{DenseTrack}, is illustrated in Fig.~\ref{fig_tracking_block_diagram}.
\subsection{Frame-by-Frame Association}

Let $\boldsymbol{F}^{t}=\{\boldsymbol{f}_{i}^{t}|i=1,2,..,m\}$ and $\boldsymbol{F}^{t+1}=\{\boldsymbol{f}_{j}^{t+1}|j=1,2,..,n\}$ denote two consecutive frames with $m$ and $n$ cell instances, respectively, where each instance is represented by a feature vector $\boldsymbol{f}$. For each instance  in frame $t$, there exist several matching candidates in frame $t+1$, represented by the set $\boldsymbol{M}_{i}=\{(\boldsymbol{f}_{i}^t,\boldsymbol{f}_{j_{k_{i}}}^{t+1})|k_{i}=1,2,...,N_{i}\}$. These candidates can be selected from the neighborhood of the projected location of $\boldsymbol{f}_{i}^t$ in frame $t+1$. Our objective is to estimate the likelihood of each of these candidates being a correct association. For any candidate $k_{i}$ association, we create a spatiotemporal feature vector, $\boldsymbol{f}_{i,(j_{k_{i}})}^{tem}=[\boldsymbol{f}_{i}^{t-r},...,\boldsymbol{f}_{i}^t,\boldsymbol{f}_{j_{k_{i}}}^{t+1}]$. This vector is formed by concatenating the feature vector at time $t$ with the feature vectors from the preceding $r$ time frames and the feature vector of the candidate at time $t+1$. Representative features to characterize $\boldsymbol{f}_{i}$ at a time point include 3D spatial coordinates and bounding box measures of the instance. 

By leveraging the near-temporal history within our spatiotemporal feature vector, we propose computing the probability that any candidate association $k_{i}$ is correct, denoted as $P[y=1|\boldsymbol{f}_{i,(j_{k_{i}})}^{tem}]$, through the execution of a temporal sequence classification task using deep learning. We have chosen InceptionTime~\cite{ismail2020inceptiontime}, a widely adopted time series convolutional neural network model based on the Inception architecture, for this classification task. By incorporating Inception modules along with residual connections, the InceptionTime architecture is designed to address overfitting and vanishing gradient concerns. In Section \ref{sec:results_ablation}, we have demonstrated that the InceptionTime architecture outperforms other state-of-the-art time series classifiers in this classification task as part of our cell tracking framework. We have pretrained the classification network to distinguish between correct and incorrect associations ($y=1$ or $0$). During tracking execution, the network's confidence score is utilized as the association score for the $k_{i}$ candidate association, denoted as $a(\boldsymbol{f}_{i}^t, \boldsymbol{f}_{j_{k_{i}}}^{t+1})=P[y=1|\boldsymbol{f}_{i,(j_{k_{i}})}^{tem}; \boldsymbol{\Theta})]$. Here, $\boldsymbol{\Theta}$ represents the learned parameters of the network.
Overall, with $N_{i}$ number of potential associations for the $i^{th}$ instance, there are a total of $N=\sum_{i=1}^{m} {N_{i}}$ possible associations between frame $t$ and $t+1$, such that $\boldsymbol{M}=\cup_{i=1}^{m}\boldsymbol{M}_{i}$ exist. The network estimates association scores for all these $N$ associations in one shot.

Next, we establish a one-to-one correspondence between frames $t$ and $t+1$ by solving a constrained optimization problem, utilizing the calculated association scores. The objective is to choose the associations from the $N$ potential associations that maximize the sum of the association scores. Mathematically, the optimal matching approach involves searching for a solution represented by a binary vector $\boldsymbol{x}_{0}=\{0,1\}^{N}$ that maximize the objective function presented in equation~\eqref{eq:tracking_1},
\begin{equation}
    \boldsymbol{x}_{0}=\argmax_{\boldsymbol{x}\in \{0,1\}^{N}} ~ \sum_{k=1}^N \left(\boldsymbol{x}(k)\, a(\boldsymbol{f}_{i_{k}}^{t},\boldsymbol{f}_{j_{k}}^{t+1}) \right)
    \label{eq:tracking_1}
\end{equation}
The matching constraint that ensures bi-directional one-to-one correspondence for the optimization in \eqref{eq:tracking_1} can be expressed as follows,
\begin{equation} 
    \boldsymbol{Y}\boldsymbol{x}\leq \boldsymbol{b}
    \label{eq:tracking_2}
\end{equation}
where $\boldsymbol{Y}$ represents a $(m+n)\times N$ dimensional system matrix and $\boldsymbol{b}$ represents a $(m+n)$ dimensional vector of ones. The system matrix $\boldsymbol{Y}$ is designed as follows,
\begin{equation}
    \boldsymbol{Y}(q,k)=\begin{cases}
    1, & \text{if $q=i_{k}$ or $j_{k}$}\\
    0, & \text{otherwise}
  \end{cases}
  \text{;} \, \quad \text{$q=1,2,.....,(m+n)$}
   \label{eq:tracking_3}
\end{equation}
The entries of the $k^{th}$ column of $\boldsymbol{Y}$ indicate which cell instances in frame $t$ and $t+1$ correspond to the $k^{th}$ possible match $(\boldsymbol{f}_{i_{k}}^{t},\boldsymbol{f}_{j_{k}}^{t+1})$ in $\boldsymbol{M}$, where $k=1,2,....,N$. The solution to the optimization problem in equation~\eqref{eq:tracking_1} is obtained following the proposed one-to-one matching algorithm presented in Algorithm~\ref{alg:one-to-one-match}. The algorithm iterates for each instance in frame $t$ to identify its matching candidate in $t+1$ with the highest association score. In cases where two instances from $t$ are matched with a single instance in $t+1$, the instance with the higher association score is considered correct, and the other instance is assigned to its candidate with the next highest association score. The algorithm continues until all the matched instances in $t+1$ are unique cell IDs. The computational complexity of the proposed matching algorithm is $\mathcal{O}(m\log n)$. 

After performing one-to-one matching between any two consecutive frames $t$ and $t+1$, the matched instances ($\boldsymbol{x}(k)=1$) in frame $t+1$ are assigned the same identification numbers or cell IDs as their corresponding instances in frame $t$. The unmatched instances ($\boldsymbol{x}(k)=0$) in frame $t+1$ are labeled with new cell IDs. 

\subsection{Cell Division Detection}
\label{sec:tracking_theory_division}
To identify division events, we examine the unmatched instances identified throughout the video sequence since these instances may result from a cell division event or indicate the appearance of a new cell in the field of view. To determine if an unmatched instance is a potential daughter cell, we propose a novel eigendecomposition-based technique. This approach can accurately account for the rod-shaped geometry of the bacterial cell, as well as the fact that division results in the parent cell splitting into daughter cells along its major axis, as illustrated in Fig.~\ref{tracking_block_diagram_3}.

Let the coordinates of an unmatched instance be denoted by $X \in \mathcal{R}^{p \times 3}$ with $p$ representing the number of 3D points. The covariance matrix can be expressed as $A=X^{T} X$, and we perform the singular-value decomposition, $[U, S, V]=svd(A)$. The Eigenvector matrix, $V=[v_{1}, v_{2}, v_{3}]$ with $v_{i}\in \mathcal{R}^{3}$ contains three principal components, each of dimension 3. We then consider a neighborhood around $X$ with a neighborhood size twice the length of $X$ and project each neighboring cell $Y_{i} \in \mathcal{R}^{q \times 3}$ onto the $2^{nd}$ and $3^{rd}$ principal components of $X$. The resulting projection matrix is expressed as $PM_{i}=Y_{i}V_{2,3}$ and a single projection value is computed as $PV_{i}=norm(mean(PM_{i}))$. Finally, the neighboring cell $Y_{i}$ with the minimum projection value, $\argmin \{PV_{i}\}$, is considered as the other candidate daughter cell of $X$, denoted by $X^{'}$. 

Now, to further ensure that instances $X$ and $X^{'}$ in any frame $t+1$ result from the cell division of a parent cell in frame $t$, we compare the volume of the other candidate daughter cell in the current frame, $vol(X_{t+1}^{'})$, against the volume of its matched instance in the preceding time frame, $vol(X_{t}^{'})$. 
Since cell division leads to the parent cell dividing into two daughter cells, each with approximately half the volume of the parent cell, we examine whether the ratio $\frac{vol(X_{t}^{'})}{vol(X_{t+1}^{'})} \approx 50\%$. If the condition is satisfied, it suggests that $X$ and $X^{'}$ are the daughters of a parent cell from the previous frame. In such cases, we assign a distinct new cell ID to the other daughter cell $X'$ to differentiate it from its parent in the previous frame.

 \subsection{Generate Complete Trajectories}
Following frame-by-frame association and cell division detection, the complete trajectories of the labeled cell instances can be computed. This process begins by identifying the unique instance IDs in the relabeled sequence. For each unique instance ID $l$, the sequence is traversed to determine the initial and final time frames at which the instance appears, denoted by $t_{init}^{l}$ and $t_{fin}^{l}$, respectively. Additionally, the coordinates of the $l^{th}$ instance at each time frame between $t_{init}^{l}$ and $t_{fin}^{l}$ are extracted and stored in a set of coordinates represented by $\boldsymbol{C}^{l}$. Furthermore, incidents of cell division ($P(l)=l'$ or $P(l)=0$) are recorded based on the findings from Section~\ref{sec:tracking_theory_division}.

\begin{algorithm}
\caption{One-to-One Matching between Frames $t$ and $t+1$}
\begin{algorithmic}[1]
\STATE \textbf{Input}: Cell IDs for all candidate associations, $\boldsymbol{C}_{N\times 2}$ ; association scores, $\boldsymbol{a}_{N\times 1}$

\STATE \textbf{Output}: Association prediction $\boldsymbol {x}_{N\times 1}\in \{0,1\}$
\STATE  $k_{i} \leftarrow \text{no. of nearest neighbors in $t+1$ for $i^{th}$ instance in $t$} $ (set to 4)
\STATE $\boldsymbol{c}_{0} \leftarrow \text{unique cell IDs from frame $t$}$
\STATE $\boldsymbol{c}_{1} \leftarrow \text{unique cell IDs from frame $t+1$}$
\STATE $\boldsymbol{x}\leftarrow zeros_{N\times 1}$ \algorithmiccomment{initialize}
\STATE $conflict \leftarrow 1$ \algorithmiccomment{initialize}
\STATE $\boldsymbol{D}_{0}\left[\boldsymbol{c}_{0}[i]\right] \leftarrow \text{-}1 \quad \forall \,i=\{0,1,..,(len(\boldsymbol{c}_{0})\text{-}1)\}$
\algorithmiccomment{initialize a dictionary for unique IDs of $t$}
\STATE $\boldsymbol{D}_{1}\left[\boldsymbol{c}_{1}[j]\right] \leftarrow \text{-}1 \quad \forall \,j=\{0,1,..,(len(\boldsymbol{c}_{1})\text{-}1)\}$
\algorithmiccomment{initialize a dictionary for unique IDs of $t+1$}
\WHILE{$conflict > 0$}
\STATE $conflict \leftarrow 0$ 
\FOR{$i=0$ to $(len(\boldsymbol{c}_{0})\text{-}1)$}
\IF{$\boldsymbol{D}_{0}\left[\boldsymbol{c}_{0}[i]\right]=\text{-}1$}
     \STATE $max\_loc \leftarrow \argmax \boldsymbol{a}_{k_{i}\times 1}$
     \algorithmiccomment{select association with max score among $k_{i}$ candidate scores}
 \IF{$\boldsymbol{D}_{1}\left[\boldsymbol{C}[max\_loc,1]\right]=\text{-}1$}
     \STATE $\boldsymbol{D}_{1}\left[\boldsymbol{C}[max\_loc,1]\right] \leftarrow max\_loc$
     \algorithmiccomment{update with new association location}
     \STATE $\boldsymbol{D}_{0}\left[\boldsymbol{C}[max\_loc,0]\right] \leftarrow max\_loc$
     \algorithmiccomment{update with new association location}
 \ELSE
     \STATE $conflict \leftarrow 1$ 
    \IF{$ \boldsymbol{a}[max\_loc] > \boldsymbol{a}\left[\boldsymbol{D}_{1}\left[\boldsymbol{C}[max\_loc,1]\right]\right]$}
    \STATE $ \boldsymbol{a}\left[\boldsymbol{D}_{1}\left[\boldsymbol{C}[max\_loc,1]\right]\right]\leftarrow 0$
     \algorithmiccomment{indicates no association}
    \STATE $\boldsymbol{D}_{0}\left[\boldsymbol{C}[\boldsymbol{D}_{1}\left[\boldsymbol{C}[max\_loc,1]\right]\right]] \leftarrow \text{-}1$
     \algorithmiccomment{indicates no association}
    \STATE $\boldsymbol{D}_{0}\left[\boldsymbol{C}[max\_loc,0]\right] \leftarrow max\_loc$
     \algorithmiccomment{update with new association location}
     \STATE $\boldsymbol{D}_{1}\left[\boldsymbol{C}[max\_loc,1]\right] \leftarrow max\_loc$
     \algorithmiccomment{update with new association location}
    \ELSE
     \STATE $\boldsymbol{a}[max\_loc]\leftarrow 0$ \algorithmiccomment{indicates no association}  
\ENDIF   
\ENDIF
\ENDIF
\ENDFOR
\ENDWHILE
\STATE $\boldsymbol{x}\left[\boldsymbol{D}_{0}\left[\boldsymbol{c}_{0}[i]\right]\right] \leftarrow 1 \quad \forall \,i=\{0,1,..,(len(\boldsymbol{c}_{0})\text{-}1)\} \, \land \boldsymbol{D}_{0}\left[\boldsymbol{c}_{0}[i]\right] \neq \text{-}1$
\algorithmiccomment{obtain final association prediction}
\end{algorithmic}
\label{alg:one-to-one-match}
\end{algorithm}
\section{Experimental Framework}
In this section, we describe the dataset, provide the implementation details, discuss the evaluation metrics, and summarize the competing approaches. 
\subsection{Dataset}
We evaluated the proposed cell tracking method on both synthetic and real 3D microscopy image sequences of bacterial biofilms. The synthetic biofilm sequences were generated using a simulation framework~\cite{toma2022realistic} which models biofilm formation following biophysical rules and represents bacterial cells with realistic curvilinear morphology. In these synthetic sequences, starting with one or multiple seed cells, the imaged biofilm continues to form as the cells grow and divide over a period of time. We simulated multiple synthetic sequences with varying number of initial clusters where the seed cells are placed at random spatial allocations and orientations. These sequences were generated at a frame interval of 10 seconds. Each synthetic video has a dimension of $450\times 450\times 150 \times 40$ voxels in $x$-$y$-$z$-$t$. The challenge here lies in linking cell instances within a highly dense environment and detecting frequent division events.

For cell tracking in real biofilm sequences, we acquired lattice light-sheet microscopy~\cite{zhang20193d} videos of two kinds of bacteria species, \textit{Escherichia coli} and \textit{Shewanella oneidensis}. The resolution of each frame in the video is approximately 230 nm in $x$ and $y$ and 370 nm in $z$, assuming green fluorescent protein (GFP) excitation and emission. 
The \textit{S. oneidensis} video was captured at a frame interval of 30 seconds over a total period of 15 minutes, while the \textit{E. coli} sequence was captured at a frame interval of 5 minutes over a period of 50 minutes. \textit{Shewanella} bacteria species exhibit motility in dense environments, making tracking individual cells over time challenging. Conversely, the \textit{E. coli} video has a lower frame rate and features frequent division events, where cells divide with changes in orientation and spatial displacement into the next frame. This presents significant challenges in the frame-by-frame association and division event detection.

\subsection{Implementation Details}
The proposed tracking method has one module that requires training: the temporal sequence classification network. The other modules are entirely solved in the online test stage. 
We trained the network using synthetic biofilm sequences. From a training sequence, we randomly sampled trajectories, represented by a trajectory feature vector $\boldsymbol{f}_{i,(j_{k_{i}})}^{tem}=[\boldsymbol{f}_{i}^{t-r},\boldsymbol{f}_{i}^{t-1},\boldsymbol{f}_{i}^t,\boldsymbol{f}_{j_{k_{i}}}^{t+1}]$, between any frame pairs $t$ and $t+1$ with corresponding association labels of correct or incorrect associations ($y=1$ or $0$). Each $\boldsymbol{f}_{i}^t$ is represented by a 9-dimensional feature vector including 3D spatial coordinates and bounding box measures. With a choice of $r=2$, we compute a 36-dimensional feature vector $\boldsymbol{f}_{i,(j_{k_{i}})}^{tem}$ for each candidate association. Additionally, we set the number of potential associations $N_{i}=4$, with $k_{i}=1,2,...,N_{i}$. We then train the network to minimize a binary cross-entropy loss, $\mathcal{L}=-\sum_{i=1}^{m} y^{(i)} \log \hat{y}^{(i)} - (1-y^{(i)}) (1-\log \hat{y}^{(i)})$, where $\hat{y}$ is the predicted association probability for $i^{th}$ trajectory. We implemented our association network exploiting the InceptionTime architecture in thetimeseriesAI (tsai) framework~\cite{tsai}. The performance of InceptionTime network on this association task has been compared against other state-of-the-art time series networks in Section \ref{sec:results_ablation}. 

We performed tracking experiments on six synthetic sequences and two real biofilm sequences. For synthetic sequences, the experiments were performed in a leave-one-out fashion; that is, the temporal sequence classification network was pretrained on five sequences, while the tracking algorithm was evaluated on the remaining sequence. For the real image sequences of two different biofilm species, the tracking algorithm was executed using a pretrained association network on the synthetic sequences.
Since the proposed method is a tracking-by-detection approach, prior to performing the tracking task, the segmentation was performed on each 3D frame of the video using the biofilm segmentation method named \textit{DeepSeeded}, as detailed in our previous work~\cite{toma2024deepseeded,biofilm_seg_code}.

\subsection{Evaluation Measures}
We evaluated the tracking performance using two already established cell tracking performance measures. Both measures are full reference, hence compares the estimated tracks from the tracking algorithm with respect to the reference tracks. One measure is called tracking accuracy or $TRA$, which is widely adopted by the Cell Tracking Challenge. 
This metric, based on representing tracks as an acyclic oriented graph~\cite{matula2015cell}, calculates the cost associated with transforming a computed graph into the reference one. The cost, referred to as $AOGM$ (Acyclic Oriented Graph Metric), is computed as $AOGM=w_{ED}ED+w_{EA}EA+w_{EC}EC$. Here, $ED$ represents the cost of adding edges (resulting from missing links), $EA$ represents the cost of deleting edges (resulting from redundant links), and $EC$ represents the cost of altering edge semantics (resulting from incorrect division detection). The weights $w$ associated with these cost terms are typically set to 1. In essence, $TRA$ provides a relative cost compared to the expense of creating the reference graph from scratch, denoted as $AOGM_{0}$. Mathematically, the $TRA$ measure is expressed as:

\begin{equation}
    TRA = 1-\frac{\min(AOGM, AOGM_{0})}{AOGM_{0}} \nonumber
\end{equation}

Additionally, we separately evaluated the cell division detection accuracy in datasets with frequent division events using a F1 score named $\textit{Division-F1}$~\cite{ulicna2021automated} represented as follows, 
\begin{equation}
    \textit{Division-F1} =\frac{2\times precision\times recall}{precision + recall} \nonumber
\end{equation}
Here, $precision=\frac{TP}{TP+FP}$ and $recall=\frac{TP}{TP+FN}$, 
where $TP$ represents the track splitting events detected within time distance $t$ ($t=\pm 1$) of ground truth ($GT$) events, $FP$ denotes the difference between total detected events and $TP$ events, and $FN$ indicates the difference between total $GT$ events and $TP$ events. Both of these quantitative metrics are computed using a publicly available repository~\cite{traccuracy}.

\subsection{Competing Approaches}
The proposed cell tracking method $\textit{DenseTrack}$ has been evaluated against four competing approaches. We selected three recent methods that have demonstrated state-of-the-art performance in Cell Tracking Challenge datasets and have publicly available implementations. One of these methods is called $\textit{Ultrack}$, which utilizes ultrametric contours, a hierarchical representation of the image boundaries, for linking detected instances between adjacent frames through a multiple hypotheses-based technique~\cite{bragantini2023large}. Another approach, referred to as the $\textit{GraphOpt}$ approach, is a graph-based cell tracking method where segmented objects are assigned to tracks by solving a model-based graph optimization problem~\cite{loffler2021graph}. Additionally, we considered a recent deep learning-based cell tracking approach named $\textit{GNN}$, which constructs cell trajectories using a graph neural network~\cite{ben2022graph}. Finally, we compared the proposed method against a biofilm-specific tracking approach~\cite{zhang2022bcm3d} known as the $\textit{NearestNbr}$ tracking method, which performs frame-by-frame association using Euclidean distance of the extracted features.

\begin{figure*}[!t]
\centering
\subfloat[$t=10 \;sec$]{\includegraphics[width=1.68in]{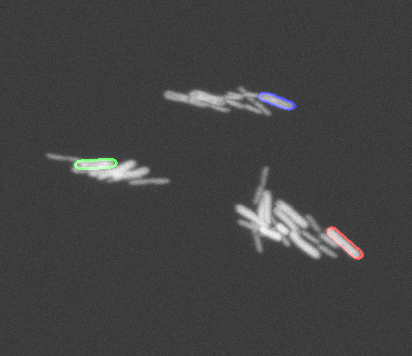}%
\label{syn_f1}}
\subfloat[$t=20 \;sec$]{\includegraphics[width=1.68in]{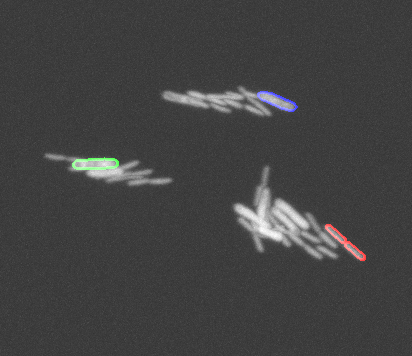}%
\label{syn_f2}}
\subfloat[$t=30 \;sec$]{\includegraphics[width=1.68in]{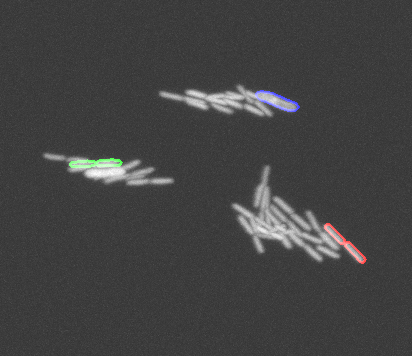}%
\label{syn_f3}}
\subfloat[$t=40 \;sec$]{\includegraphics[width=1.68in]{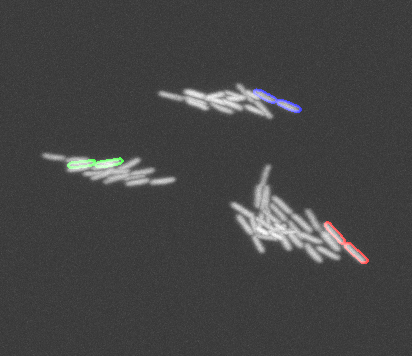}%
\label{syn_f4}}
\\
\subfloat[$t=190 \;sec$]{\includegraphics[width=1.68in]{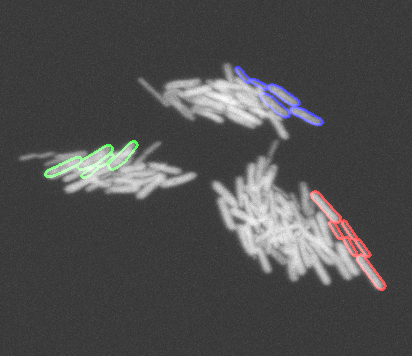}%
\label{syn_f5}}
\subfloat[$t=200 \;sec$]{\includegraphics[width=1.68in]{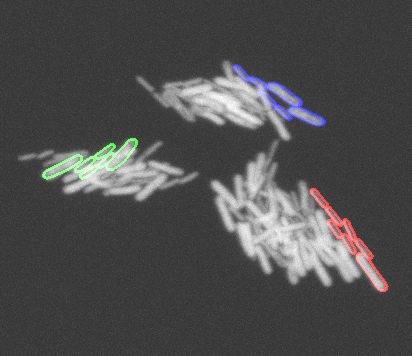}%
\label{syn_f6}}
\subfloat[$t=210 \;sec$]{\includegraphics[width=1.68in]{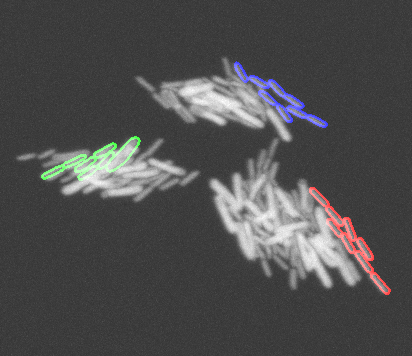}%
\label{syn_f7}}
\subfloat[$t=220 \;sec$]{\includegraphics[width=1.68in]{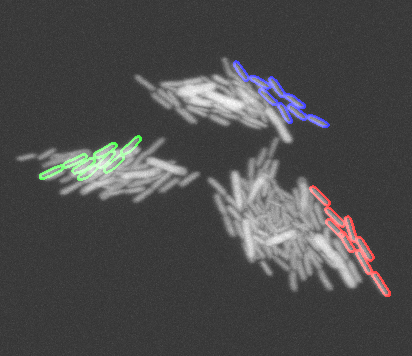}%
\label{syn_f8}}
\\
\subfloat[$t=420 \;sec$]{\includegraphics[width=1.68in]{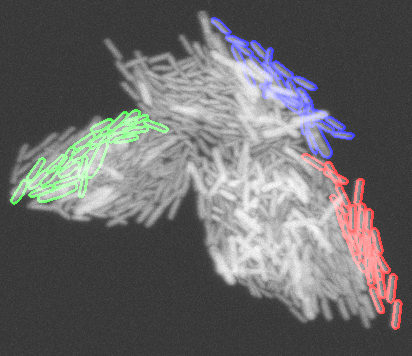}%
\label{syn_f9}}
\subfloat[$t=440 \;sec$]{\includegraphics[width=1.68in]{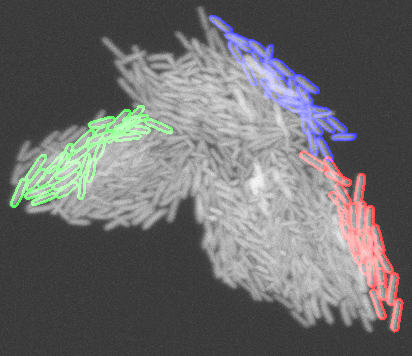}%
\label{syn_f10}}
\subfloat[$t=460 \;sec$]{\includegraphics[width=1.68in]{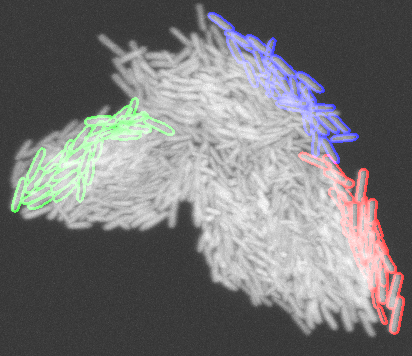}%
\label{syn_f11}}
\subfloat[$t=500 \;sec$]{\includegraphics[width=1.68in]{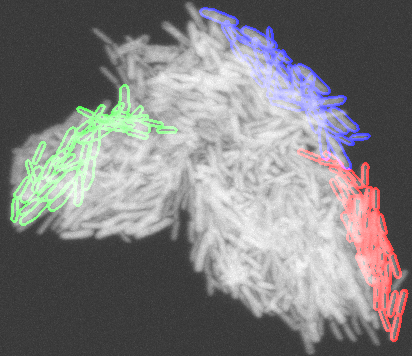}%
\label{syn_f12}}
\caption{Qualitative visualization of cell tracking by \textit{DenseTrack} in a synthetic biofilm sequence with 50 frames captured at 10 seconds frame interval. We demonstrate tracking of three particular cells at several frames in the sequence. Each 3D frame is displayed as a maximum intensity projection along z axis.}
\label{fig_syn_track_qual}
\end{figure*}

\begin{figure*}[!t]
\centering
\subfloat[]{\includegraphics[width=2.95in]{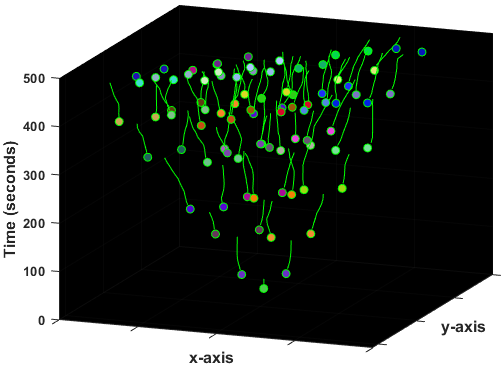}%
\label{syn_space_time}}
\hfil
\subfloat[]{\includegraphics[width=3.1in]{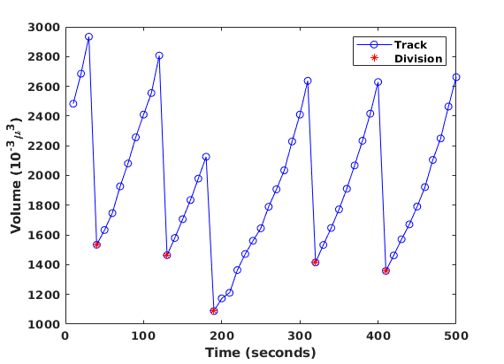}%
\label{syn_vol_over_time}}
\caption{Showing evidence of effective cell division detection over time by the \textit{DenseTrack} method through (a) space-time plot and (b) volume-over-time plot, demonstrated for the 'blue' cell in the synthetic sequence in Fig.~\ref{fig_syn_track_qual}.}
\label{fig_syn_track_spacetime_vol}
\end{figure*}

\begin{table}[!tbp]
\centering
\setlength{\tabcolsep}{10.2pt}
\renewcommand{\arraystretch}{1.5}
\caption{Quantitative tracking evaluation on six synthetic biofilm videos\label{tab:quant_synthetic}}
\begin{tabular}{|c||c|c|}
\hline
\textbf{Methods} & \textbf{TRA} & \textbf{Division-F1} \\\hline
\textit{DenseTrack} & \textbf{0.942 $\pm$ 0.018} & \textbf{0.911 $\pm$ 0.022}\\\cline{1-3}
\textit{Ultrack}~\cite{bragantini2023large} & 0.919 $\pm$ 0.021 & 0.864 $\pm$ 0.024\\\cline{1-3}
\textit{GraphOpt}~\cite{loffler2021graph} & 0.915 $\pm$ 0.019 & 0.886 $\pm$ 0.022\\\cline{1-3}
\textit{NearestNbr}~\cite{zhang2022bcm3d} & 0.840 $\pm$ 0.022& 0.648$\pm$ 0.025\\\cline{1-3}
\textit{GNN}~\cite{ben2022graph} &0.818 $\pm$ 0.026 & 0.637 $\pm$ 0.033\\\cline{1-3}
\end{tabular}
\end{table}

\begin{figure*}[!t]
\centering
\subfloat[$t=0.5 \;min$]{\includegraphics[width=1.6in]{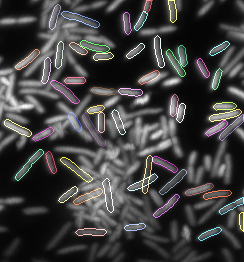}%
\label{shewn_f1}}
\subfloat[$t=2.5 \;min$]{\includegraphics[width=1.6in]{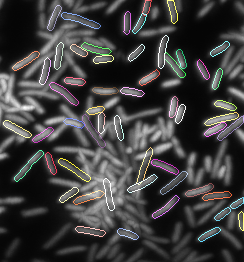}%
\label{shewn_f2}}
\subfloat[$t=5 \;min$]{\includegraphics[width=1.6in]{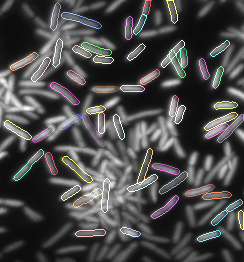}%
\label{shewn_f3}}
\subfloat[$t=7.5 \;min$]{\includegraphics[width=1.6in]{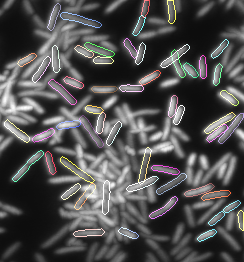}%
\label{shewn_f4}}
\\
\subfloat[$t=10 \;min$]{\includegraphics[width=1.6in]{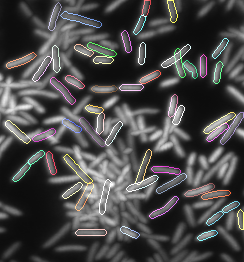}%
\label{shewn_f5}}
\subfloat[$t=12.5 \;min$]{\includegraphics[width=1.6in]{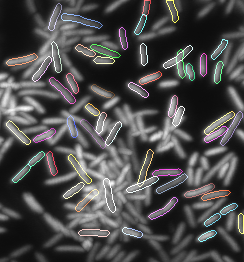}%
\label{shewn_f6}}
\subfloat[$t=13.5 \;min$]{\includegraphics[width=1.6in]{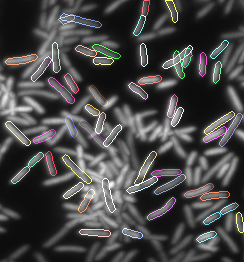}%
\label{shewn_f7}}
\subfloat[$t=15 \;min$]{\includegraphics[width=1.6in]{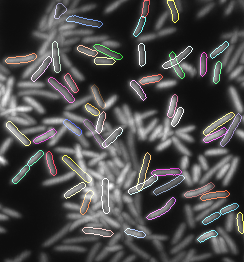}%
\label{shewn_f8}}
\caption{Qualitative observation of bacterial cell tracking using the \textit{DenseTrack} method in a real \textit{S. oneidensis} biofilm sequence, consisting of 30 frames captured at 30-second intervals. We display the predicted matched instances for a group of randomly selected cells over various time points in the video, each represented by a distinct color.}
\label{fig_shewn_track_qual}
\end{figure*}

\begin{figure*}[!t]
\centering
\subfloat[]{\includegraphics[width=2.959in]{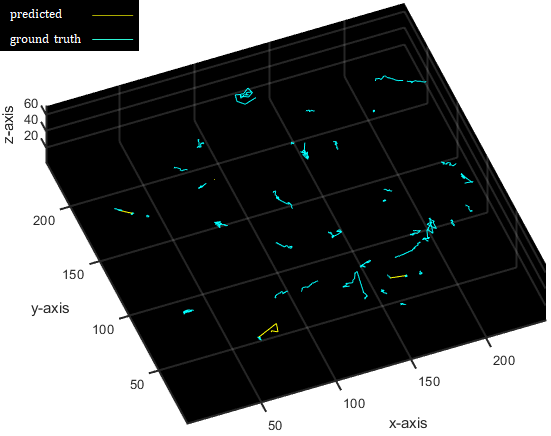}%
\label{shewn_space_trajectory_plot_proposed}}
\hfil 
\subfloat[]{\includegraphics[width=3.0in]{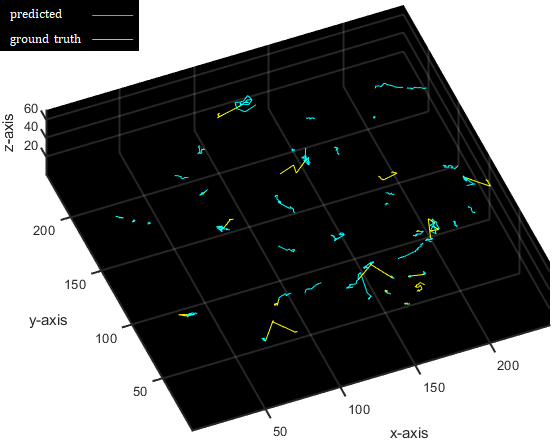}%
\label{shewn_space_trajectory_plot_ultrack}}
\caption{Visualizing thirty predicted trajectories of the \textit{S. oneidensis} sequence obtained from (a) the $\textit{DenseTrack}$ method and (b) the $\textit{Ultrack}$ method, in comparison to the corresponding manually labeled ground truth trajectories. The spatial dimension is $244\times 262\times 87$ voxels in $x$-$y$-$z$. The trajectories depicted in Fig.\ref{shewn_space_trajectory_plot_proposed} exhibit greater alignment with the ground truth.}
\label{fig_shewn_trajectory_compare}
\end{figure*}

\begin{figure}[!t]
\centering
\includegraphics[width=3.45in]{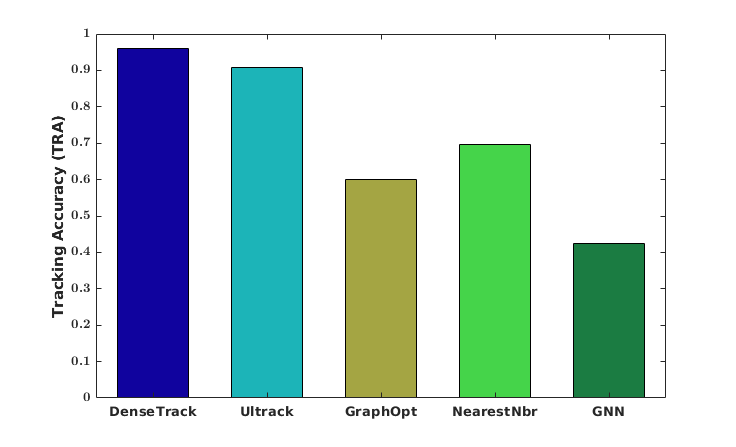}
\caption{Quantitative tracking evaluation on \textit{S. oneidensis} video in Fig.~\ref{fig_shewn_track_qual}.}
\label{fig_shewn_quant}
\end{figure}

\begin{figure*}[!t]
\centering
\subfloat[$t=5 \;min$]{\includegraphics[width=1.8in]{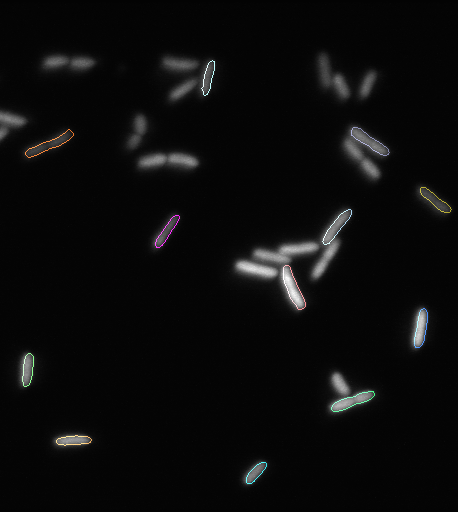}%
\label{ecoli_f1}}
\subfloat[$t=20 \;min$]{\includegraphics[width=1.8in]{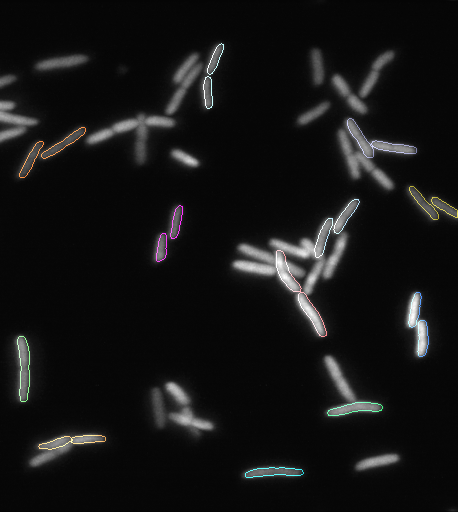}%
\label{ecoli_f2}}
\subfloat[$t=25 \;min$]{\includegraphics[width=1.8in]{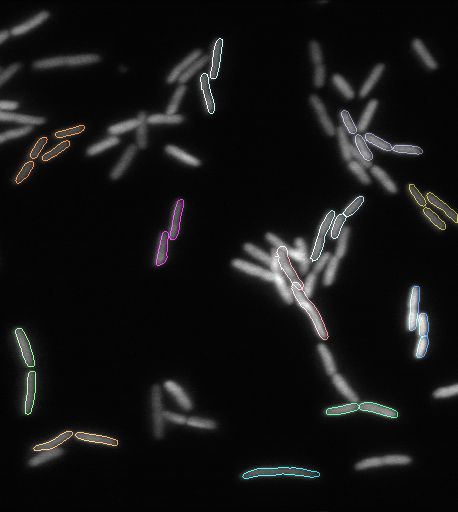}%
\label{ecoli_f3}}
\\
\subfloat[$t=35 \;min$]{\includegraphics[width=1.8in]{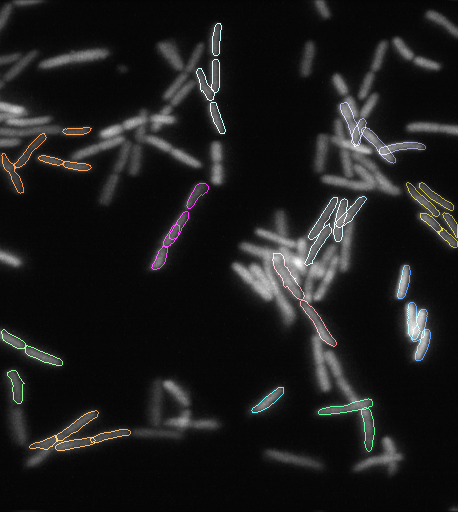}%
\label{ecoli_f4}}
\subfloat[$t=45 \;min$]{\includegraphics[width=1.8in]{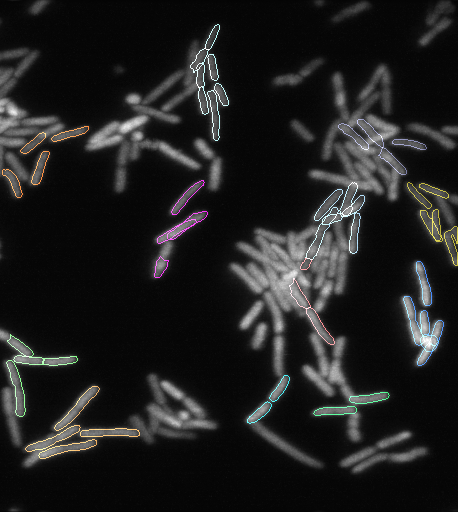}%
\label{ecoli_f5}}
\subfloat[$t=50 \;min$]{\includegraphics[width=1.8in]{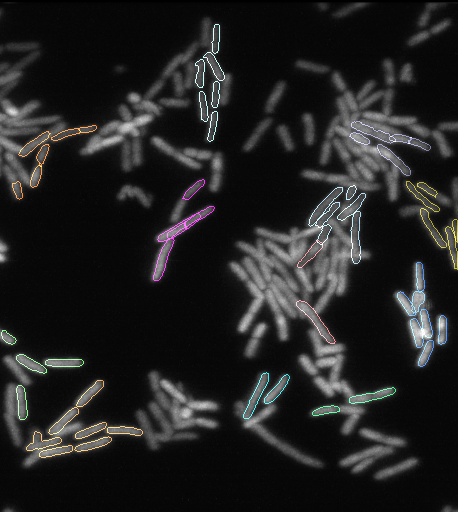}%
\label{ecoli_f6}}
\caption{Visualization of cell tracking results obtained using the \textit{DenseTrack} method in an \textit{E. coli} biofilm video consisting of 10 frames captured at five-minute intervals. The figure displays the tracked instances for randomly selected cells across different time points in the video, each in a unique color.}
\label{fig_ecoli_track_qual}
\end{figure*}

\begin{figure*}[!t]
\centering
\subfloat[]{\includegraphics[width=2.959in]{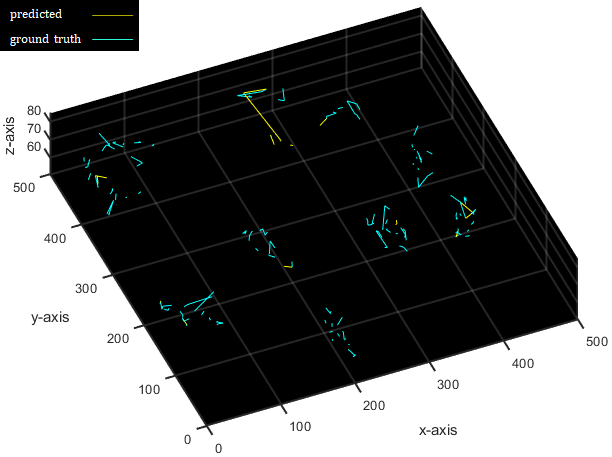}%
\label{ecoli_space_trajectory_plot_proposed}}
\hfil 
\subfloat[]{\includegraphics[width=2.959in]{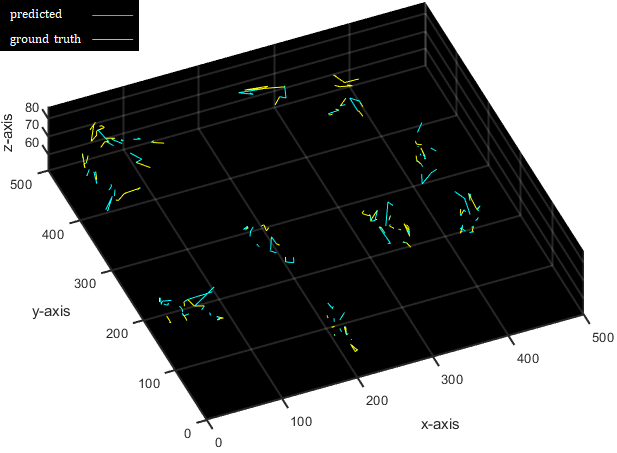}%
\label{ecoli_space_trajectory_plot_ultrack}}
\caption{Comparing ten predicted trajectories of the \textit{E. coli} sequence from (a) the $\textit{DenseTrack}$ method and (b) the $\textit{Ultrack}$ method to the corresponding manually labeled ground truth trajectories. The spatial dimension is $458\times 512\times 101$ voxels in $x$-$y$-$z$. The $\textit{DenseTrack}$ method shows more overlap with the ground truth. }
\label{fig_ecoli_trajectory_compare}
\end{figure*}

\begin{table}[!tbp]
\centering
\setlength{\tabcolsep}{10.5pt}
\renewcommand{\arraystretch}{1.5}
\caption{Quantitative tracking evaluation on an \textit{E. coli} biofilm video\label{tab:quant_ecoli}}
\begin{tabular}{|c||c|c|}
\hline
\textbf{Methods} & \textbf{TRA} & \textbf{Division-F1} \\\hline
\textit{DenseTrack} & \textbf{0.904} & \textbf{0.877}\\\cline{1-3}
\textit{Ultrack}~\cite{bragantini2023large} & 0.823  & 0.652\\\cline{1-3}
\textit{GraphOpt}~\cite{loffler2021graph} &0.764 & 0.410\\\cline{1-3}
\textit{NearestNbr}\cite{zhang2022bcm3d} & 0.512 & 0.391\\\cline{1-3}
\textit{GNN}~\cite{ben2022graph} & 0.477& 0.297\\\cline{1-3}
\end{tabular}
\end{table}

\begin{table}[!tbp]
\centering
\setlength{\tabcolsep}{10.5pt}
\renewcommand{\arraystretch}{1.5}
\caption{Binary classification accuracy on temporal sequence classification using various classifiers, and using classifier's confidence scores in a one-to-one matching ($OTOM$) optimization
\vspace{0.5em}
\label{tab:ablation_classifiers_tracking}}
\begin{tabular}{|c||c|c|}
\hline
\textbf{Methods} & \textbf{Classifier} & \textbf{Classifier+OTOM} \\\hline
\textit{InceptionTime}~\cite{ismail2020inceptiontime} & \textbf{0.964 $\pm$ 0.007} & \textbf{0.998 $\pm$ 0.003}\\\cline{1-3}
\textit{TST}~\cite{zerveas2021transformer} & 0.886 $\pm$ 0.032 & 0.940 $\pm$ 0.013\\\cline{1-3}
\textit{LSTM-FCN}~\cite{karim2017lstm} & 0.914 $\pm$ 0.024 & 0.958 $\pm$ 0.006\\\cline{1-3}
\textit{GRU-FCN}~\cite{elsayed2018deep} & 0.919 $\pm$ 0.018 & 0.952$\pm$ 0.007\\\cline{1-3}
\textit{Res-CNN}~\cite{zou2019integration} & 0.804 $\pm$ 0.031 & 0.909 $\pm$ 0.009\\\cline{1-3}
\end{tabular}
\end{table}

\begin{figure}[!tb]
\centering
\subfloat[]{\includegraphics[width=1.7in]{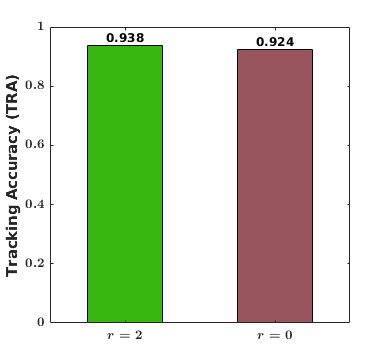}%
\label{ablation_synthetic}}
\subfloat[]{\includegraphics[width=1.7in]{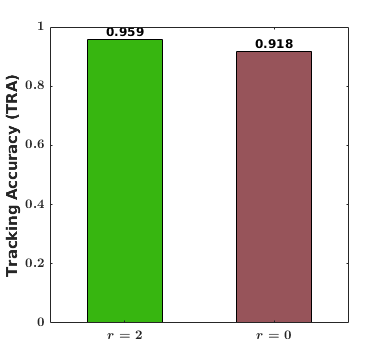}%
\label{ablation_shewn}}
\caption{Evidence of exploiting near-temporal history ($r=2$) in tracking performance, on a (a) synthetic biofilm video, and a (b) real biofilm video of \textit{S. oneidensis}.}
\label{fig_ablation_tracking}
\end{figure}

\section{Results and Discussion}
\label{sec:results}
In this section, we present both qualitative and quantitative tracking results obtained from synthetic and real biofilm image sequences. In Fig.~\ref{fig_syn_track_qual}, we display the tracking results of a synthetic biofilm sequence obtained from the proposed $\textit{DenseTrack}$ algorithm. The algorithm performs the tracking task for all cell instances. However, for clarity of visual observation in a dense environment, we demonstrate the predicted matched instances of three particular cells over the length of the image sequence, displayed in red, blue, and green. The figure shows that the $\textit{DenseTrack}$ method can successfully associate the same instance of a cell over consecutive time points, even in such a crowded neighborhood. Furthermore, the cell division events are also accurately detected by the proposed method, which is essential for an effective tracking outcome in such a dataset involving frequent division events. 

In Fig.~\ref{fig_syn_track_spacetime_vol}, we exhibit additional support for the effectiveness of the proposed method in cell division detection using a space-time plot and a volume-over-time plot. We demonstrate these two plots for the `blue' cell of the displayed sequence in Fig.~\ref{fig_syn_track_qual}. The space-time plot depicts the $x$ and $y$ coordinates of the centroid of the 'blue' cell and its matched instances over time. The green circle at the bottom represents the cell's location in the first frame. Pairs of circles in the same color indicate that the tracking algorithm detects two daughter cells in that space and time. The line growing out of the circle signifies the instance's growth until it divides again. Additionally, the space-time plot effectively demonstrates that the bacterial cell's division follows a geometric progression, such as 2, 4, 8, 16, and so forth. Besides, we generate the volume-over-time plot as shown in Fig.~\ref{syn_vol_over_time} by considering the volume of only one daughter cell at each division event along the sequence. The sawtooth pattern of the plot ensures that the cell divisions are correctly detected by the tracking method, as the volume increases when the cell grows and decreases as it splits into daughter cells.

In Table~\ref{tab:quant_synthetic}, we report the quantitative tracking performance for six synthetic biofilm image sequences in our dataset. These videos contain an average of 1400 ground truth division events. The comparison of tracking methods is based on the overall tracking accuracy ($TRA$) and the division-specific accuracy metric ($\textit{Division-F1}$). The results indicate that the proposed $\textit{DenseTrack}$ method outperforms other methods in both performance measures. Additionally, $\textit{Ultrack}$ and $\textit{GraphOpt}$ exhibit reasonable performance in tracking bacterial cells within a dense biofilm environment. However, the nearest neighbor-based technique $\textit{NearestNbr}$, employing a simplistic Euclidean distance-based frame-by-frame matching, and the graph neural network-based approach $\textit{GNN}$, predicting one directed graph for the entire sequence, exhibit lower tracking accuracy in both measures.

Next, we present the visualization of tracked cell instances by $\textit{DenseTrack}$ method for a real biofilm sequence of \textit{S. oneidensis} bacteral species in Fig.~\ref{fig_shewn_track_qual}, which was captured at a 30-second frame interval. The matched instances of the same cell over time are displayed in the same color. The figure demonstrates that the proposed method accurately tracks most cell instances. Furthermore, we visualize the predicted trajectories compared to corresponding ground truth trajectories in Fig.~\ref{fig_shewn_trajectory_compare}. Thirty manually generated ground truth trajectories are plotted in $x$-$y$-$z$ on top of the estimated trajectories by the tracking algorithm. We compare such trajectory plots from the proposed $\textit{DenseTrack}$ method and the best-competing method, $\textit{Ultrack}$, as shown in Fig.~\ref{shewn_space_trajectory_plot_proposed} and \ref{shewn_space_trajectory_plot_ultrack}. 
Observing the figures, it becomes apparent that predicted trajectories from $\textit{DenseTrack}$ demonstrate a higher degree of overlap with the ground truth, suggesting superior accuracy compared to the $\textit{Ultrack}$ method.

In Fig.~\ref{fig_shewn_quant}, we further present a comparative analysis of quantitative tracking performance based on the aforementioned thirty ground truth trajectories. Since the \textit{S. oneidensis} sequence exhibits very few cell division events (only three ground truth division events in the thirty trajectories), we have opted not to separately present the $\textit{Division-F1}$ measure and instead focus on reporting the overall tracking score $TRA$. The figure reveals that, similar to the results obtained from synthetic videos, the proposed $\textit{DenseTrack}$ approach excels in tracking motile \textit{S. oneidensis} bacterial cells. While the $\textit{Ultrack}$ method also demonstrates reasonable performance with approximately 90\% tracking accuracy, 
the $\textit{GraphOpt}$, $\textit{NearestNbr}$ and $\textit{GNN}$ methods encounter challenges tracking instances within this real biofilm sequence, leading to lower $TRA$ scores.

We then showcase the qualitative tracking results of our proposed approach on an \textit{E. coli} image sequence in Fig.~\ref{fig_ecoli_track_qual}, captured at a larger frame interval of 5 minutes. In this figure, we observe that even in a lower frame-rate video with very frequent division events, the proposed method performs reasonably well in tracking the bacteria cells and their offspring. Furthermore, we offer a qualitative comparison of spatial trajectory plots between the proposed method and the $\textit{Ultrack}$ method with respect to ten manually generated ground truth trajectories in Fig.~\ref{fig_ecoli_trajectory_compare}. While in comparison to the trajectory plot for the \textit{S. oneidensis} sequence (Fig.~\ref{shewn_space_trajectory_plot_proposed}), the proposed method exhibits more deviations from ground truth for this \textit{E. coli} sequence (Fig.~\ref{ecoli_space_trajectory_plot_proposed}), such deviations are still fewer than those obtained from the $\textit{Ultrack}$ method (Fig.~\ref{ecoli_space_trajectory_plot_ultrack}).

In Table~\ref{tab:quant_ecoli}, we present the $TRA$ and \textit{Division-F1} scores of the comparative methods based on the ten manually generated trajectories depicted in Fig.~\ref{fig_ecoli_trajectory_compare}. These trajectories cover a total of 54 ground truth division events. The table illustrates that our proposed $\textit{DenseTrack}$ approach achieves superior tracking performance even in this lower frame rate video (5-minute interval). Additionally, it is observed that the performances of the four competing methods deteriorate further, resulting in lower \textit{Division-F1} scores compared to their performance on synthetic image sequences in Table~\ref{tab:quant_synthetic}. This decrease in performance in division prediction for the \textit{E. coli} sequence may be attributed to the presence of complex division events accompanied by orientation changes and spatial displacement in the next frame.

\subsection{Ablation Study}
\label{sec:results_ablation}
To comprehend the distinct contributions of various components in our proposed method, we conduct ablation studies and report the results in this section. In Table~\ref{tab:ablation_classifiers_tracking}, we present quantitative support for our selection of the \textit{InceptionTime} classifier in the temporal sequence classification task for frame-by-frame association.
The first column presents the classification accuracy of different classifiers in distinguishing correct from wrong associations in temporal sequences. We determine classification accuracy based on whether the confidence score for the `correct' class exceeds that of the `wrong' class for a given association. Among the classifiers, \textit{InceptionTime} demonstrates superior performance. However, classification in the first column may contain errors stemming from incorrect mappings between frames, such as one-to-multiple associations. The second column reveals that integrating the classifier's confidence scores into one-to-one matching optimization, as in our \textit{DenseTrack} framework, enhances classification performance across all classifiers. Nevertheless, \textit{InceptionTime} still achieves the best results. These classification scores are averaged over 30 frame pairs from two synthetic biofilm videos, each with 15 frames randomly selected.

In Fig.~\ref{fig_ablation_tracking}, we highlight the importance of leveraging near-temporal history in the temporal sequence classification task as part of our proposed tracking approach, rather than solely relying on cellular attributes from the present frame and the next frame. The significance is measured in terms of the overall tracking accuracy measure $TRA$. In Section \ref{sec:tracking_theory}, we mentioned the use of a spatiotemporal feature vector, $\boldsymbol{f}_{i,(j_{k_{i}})}^{tem}=[\boldsymbol{f}_{i}^{t-r},...,\boldsymbol{f}_{i}^t,\boldsymbol{f}_{j_{k_{i}}}^{t+1}]$, formed by concatenating the feature vector at time $t$ with the feature vectors from the preceding $r$ time frames and the feature vector at time $t+1$. The figure illustrates the effect of using $r=2$ as in our proposed method versus the effect of using $r=0$. In Fig.~\ref{ablation_synthetic}, we observe such a comparison for a synthetic biofilm video, while in Fig.~\ref{ablation_shewn}, we observe it for a \textit{S. oneidensis} video. The figures indicate that utilizing near-temporal history ($r=2$) improves tracking accuracy for both the synthetic sequence and the real biofilm sequence, with a more pronounced improvement observed in the real biofilm example.

\section{Conclusion}

This paper introduced a novel cell tracking approach to effectively track cell instances and their offspring in dense 3D time-lapse microscopy image sequences. We formulated the cell tracking problem as a frame-by-frame matching task exploiting a deep temporal sequence classifier's confidence scores in a one-to-one optimization framework. Utilizing a data-driven deep-learning-based classifier as opposed to a fixed distance or similarity-based measure resulted in better association scores for the potential matches between frame pairs. Additionally, an effective one-to-one matching optimization formulation with proper constraints presented in this work ensures superior performance in associating cell instances within a crowded environment. To detect cell division events with high accuracy, we also proposed an eigendecomposition-based strategy that can identify division events even when daughter instances change orientation and displace spatially during dividing from the parent instance. We demonstrate the effectiveness of the proposed method in tracking bacterial cells from 3D lattice light-sheet image sequences of biofilms. The proposed method achieved better results than the state-of-the-art cell tracking approaches.

\section*{Acknowledgments}
This work is supported in part by the U.S. National Institute of General Medical Sciences under NIH Grant No.~1R01GM139002. For this study, no ethical approval was required. The authors have no conflicts of interest.

\bibliographystyle{IEEEtran}
\bibliography{ref_tracking}

\end{document}